\documentclass[aps,preprint,prd,epsfig,axodraw,nofootinbib]{revtex4-1}
\usepackage{amssymb}
\usepackage{amsmath}
\usepackage{graphicx}
\usepackage{fancyhdr}
\usepackage{epsfig}
\usepackage{axodraw}
\def\br{\begin{eqnarray}}
\def\er{\end{eqnarray}}
\def\be{\begin{equation}}
\def\ee{\end{equation}}

\def\g{\gamma}

\def\l{\lambda}

\def\to{\rightarrow}

\def\({\left(}
\def\){\right)}

\def\invfb{\; \hbox{fb}^{-1}}
\def\gev{\; \hbox{GeV}}

\def\lesssim{\mathrel{\hbox{\rlap{\hbox{\lower4pt\hbox{$\sim$}}}\hbox{$<$}}}}
\def\gtrsim{\mathrel{\hbox{\rlap{\hbox{\lower4pt\hbox{$\sim$}}}\hbox{$>$}}}}
\begin{document}

\title{Probing 3-3-1 Models in Diphoton Higgs Boson Decay}

\author{A. Alves}
\affiliation{{Departamento de Ci\^encias Exatas e da Terra,
Universidade Federal de S\~ao Paulo,  Diadema-SP, 09972-270, Brasil.}}

\author{E. Ramirez Barreto, A. G. Dias}
\affiliation{{ Centro de Ci\^encias Naturais e Humanas, Universidade Federal do ABC,  Santo Andr\'e-SP, 09210-170, Brasil.}}

\author{C. A. de S. Pires$^a$, F. S. Queiroz$^{a,b}$, and P. S. Rodrigues da Silva$^{a}$}
\affiliation{{$^a$Departamento de
F\'{\i}sica, Universidade Federal da Para\'\i ba, Caixa Postal 5008, 58051-970,
Jo\~ao Pessoa, PB, Brasil\vspace{0.2 cm}\\
$^b$Center for Particle Astrophysics, Fermi National Accelerator Laboratory, Batavia, IL 60510, USA
}}

\date{\today}

\begin{abstract}
We investigate the Higgs boson production through gluon fusion and its decay into two photons at the LHC in the context of the minimal 3-3-1 model and its alternative version with exotic leptons. The diphoton Higgs decay channel presents an enhanced signal in this model compared to the Standard Model due to the presence of an extra singly charged vector boson and a doubly charged one. Prospects for the Higgs boson detection at 7~TeV center-of-mass energy with up to $10$~fb$^{-1}$ are presented. Our results suggest that a Higgs boson from these 3-3-1 models can potentially explain the small excess for $m_H\leq 145$~GeV observed at the LHC.  Otherwise, if this excess reveals to be only a statistical fluctuation of the Standard Model backgrounds severe constraints  can be put on these models.
%\\
%PACS: 14.60.St; 14.60.Pq; 12.60.Cn; 12.60.Fr.
%
\end{abstract}

\maketitle

\section{ Introduction}
\label{intro}
The search for the Higgs boson is one of the most exciting endeavors that takes place at the Large Hadron Collider (LHC). Whatever its nature, the Higgs resonant production can be more easily studied in the diphoton channel ($H\rightarrow \g\g$) if its mass is less than about 160~GeV. This is so even if its branching ratio is mild, as is the standard model (SM) case, roughly $10^{-3}$ for a Higgs mass $M_H\approx 160$~GeV~\cite{PDG}. The most recent experimental results on the search for a Higgs boson resonance through diphoton decay have been made available by the ATLAS collaboration~\cite{ATLAS2011,ATLASnew} at CERN LHC in $pp$ collisions at $\sqrt{s}=7$~TeV, and also by the CDF and D$0\!\!\!/$ collaborations~\cite{Tevatron} at Fermilab Tevatron in $p\bar{p}$ collisions at $\sqrt{s}=1.96$~TeV. Also, the CMS Collaboration~\cite{CMS} has just delivered a huge analysis of their Higgs search for a large range of masses, and a small excess of events is being observed in region below $145$~GeV.

The diphoton Higgs decay channel also provides an excellent opportunity to test some alternatives to the SM that could be realized at the electroweak scale, and hence, be promptly probed at the current phase of LHC.
Since the main Higgs production channel is through gluon fusion, there are some possibilities of augmenting the diphoton signal in some extensions of the SM by addition of heavy colored fields, charged under $SU(2)_L$, and/or increase the number of heavy electrically charged vector bosons that couple to the Higgs, strengthening  the $H\rightarrow \g\g$ amplitude~\cite{Ellis,Okun,Hunters}. Some explicit examples of such SM extensions are models with
 more fermionic generations~\cite{4thf}\footnote{It is remarkable though that a fourth SM family is already excluded by CMS for a Higgs boson mass in the range $120-600$~GeV with $95\%$~C.L.~\cite{CMS}.}, composite Higgs models~\cite{composite}, spatial extra-dimension models~\cite{gfED}, new colored scalars~\cite{Scolored}, Higgs impostors~\cite{Lykken}, supersymmetric models~\cite{hsusy}, Two Higgs Doublets Models (2HDM-type I) and Triplet Models~\cite{aalves}.

In this work we explore a class of models which are based on the gauge group  $SU(3)_C\otimes SU(3)_L\otimes U(1)_X$, and known as 3-3-1 models~\cite{earlymodels,ppf}. The models are modest in their departure from the SM, but have many new features that make them competitive with more daring proposals. Many studies concerning signals of the new  particles predicted by 3-3-1 models at current colliders have been done recently~\cite{collider331}.

It is on the peculiar particle content of the 3-3-1 models that we concentrate to look for a clear signal of its lightest scalar particle, which is also contained in a doublet under the $SU(2)_L$ gauge symmetry, and is identified as the Higgs boson. Since we are interested in achieving some enhancement on the $H\rightarrow \g\g$ signal, the best choices among the variety of 3-3-1 model versions are the minimal 3-3-1 model~\cite{ppf} and its cousin that contains an exotic lepton in place of the right-handed components of the usual charged leptons, we call it exotic lepton 3-3-1 model (EL331)~\cite{EL331}. The reason is that there are four new electrically charged vector bosons in their spectra that add to the standard $W^\pm$ contributions in the loop to the $H\rightarrow \g\g$ amplitude, one of them being doubly charged which further enhances the decay rate. As we shall see,  the new vector bosons can have a large impact on the process in question, which also offers a way to test a mass interval for those particles if the Higgs is light enough.

We analyze the window on light Higgs mass, $100 \leq m_H \leq 150$~GeV, and our results show that if the $pp\rightarrow\gamma\gamma+X$ excess is really confirmed in the next months at the CERN LHC, then the 3-3-1 models with relatively light new vector bosons become a possible explanation for this beyond the SM physics signal. On the other hand, if the background hypothesis is confirmed only very heavy new vector bosons will be consistent with the experimental results. Thus, the LHC will be able to place the most stringent constraints on the 3-3-1 models to date, in the light Higgs mass range, ruling out portions of the model's parameter space. Moreover, since these models do not bring any new contribution to the gluon fusion Higgs production and our Higgs boson decays into SM particles mostly as in the SM case, LHC can put important limits on these 3-3-1 models on similar grounds as for the SM, considering all assessable Higgs boson decay modes. The simulations were based on partonic cross sections and NLO corrections were taken into account in the computation of all backgrounds and signals.

In the following, we define the essential aspects of the models for describing the Higgs decay into two photons,  the phenomenological analysis and our results.

\section{The models}
\label{sec1}

The content of quarks and vector bosons is equal for the models we deal with~\cite{truly,EL331}. Thus, we start defining the quark sector. Quarks form the following representations, where the number inside parenthesis means transformation properties under $SU(3)_L$ and $U(1)_X$, respectively, (we are omitting $SU(3)_C$ color). Left-handed quarks compose the multiplets $Q_{1_L}^T= \left (u_1 \,,  d_1 \,, J_1\right )_L^T \sim (\mbox{{\bf 3}},\frac{2}{3}),\,$ $Q_{n_L}^T=\left (d_n\,,-u_n \,, j_n \right )^T_L \sim (\mbox{${\bf 3^*}$},-\frac{1}{3}),$ where the indices $1$ and $n=2,3$ label the quark families, with the corresponding right-handed quarks as $u_{i_R}\sim(\mbox{{\bf 1}},\frac{2}{3}),$ $d_{i_R}\sim(\mbox{{\bf 1}},-\frac{1}{3})$, $J_{1_R}\sim(\mbox{{\bf 1}},\frac{5}{3})$, and $j_{n_R}\sim(\mbox{{\bf 1}},-\frac{4}{3})$  where $i=1,2,3$. $u_{i}$, and $d_{i}$ are the standard quarks, $J_{1}$, and $j_{n_R}$  the exotic ones with electric charge $\frac{5}{3}$ and $-\frac{4}{3}$, respectively. Observe that we are omitting any superscript that would indicate that our fermions are not in a mass eigenstate basis. However, for our purposes this will be irrelevant, except that mixing would allow for some diverse phenomenological consequences not to be explored in this work.

Leptons do not play an essential role in our analysis here but we comment that two constructions are possible. In the minimal 3-3-1 model, the standard  lepton fields compose three triplets in the form $f_{l_L}^T = \left(\nu_l\,\, e_l\,\, e^{c}_l\right)^T_L\sim(1\,,\,3\,,\,0)$, where $l=e,\,\mu,\,\tau$, and the third component $e^{c}_l$ is the charge conjugation of the charged lepton field $e_l$ \cite{ppf}. In order to generate mass to all leptons a scalar sextet would be needed~\cite{ppr}. But we disregard such a scalar sextet by taking into account that in this version lepton masses can arise through nonrenormalizable effective operators~\cite{truly}. Another possible construction for the leptons sector, which characterizes the EL331 model \cite{EL331}, is such that it is composed by three leptonic triplets $f_{l_L}^T = \left (\nu_l\,\,e_l\,\,E_l^C \right )^T_L\sim(1\,,\,3\,,\,0)$, and the right-handed singlets $e_{lR}\sim(1\,,\,1\,,\,-1)$, $(E_{l}^C)_R\sim(1\,,\,1\,,\,1)$. Dirac masses can be obtained for charged leptons in the EL331 model by means of renormalizable operators \cite{EL331}\footnote{We could also add a singlet right handed neutrino to generate Dirac neutrino mass or we could rely on a effective dimension-6 operator like $\frac{h_{ij}^\nu}{\Lambda^2}(\bar f^C_{iL} \eta^*)(\eta^\dagger f_{jL}) + H.c.$ to yield a Majorana neutrino mass.}.

The gauge symmetry structure $SU(3)_L\otimes U(1)_X$ of the models implies the existence of nine vector bosons. The four standard ones, $\g$, $Z$ and $W^\pm$, and five new ones, namely, a neutral $Z^{\prime}$, two simply charged vector bosons $V^{\pm}$, and two doubly charged vector  bosons $U^{\pm \pm}$. These new charged vector  bosons affect in an important way the contribution to the Higgs width for its decay into two photons as we shall see in the next section.

The spontaneous symmetry breaking leading to the mass generation for the vector  bosons and quarks is engendered by a set of three scalar triplets, $\chi^T =\left (\chi^-,\,\chi^{--},\, \chi^0\right )^T\sim$~({\bf 3},-1), $\rho^T =\left (\rho^+ ,\, \rho^0 ,\,\rho^{++}\right )^T\sim$~ ({\bf 3},1), and $\eta^T = \left (\eta^0,\, \eta^-,\, \eta^+ \right )^T\sim$~({\bf 3}, 0). In order to correctly generate the spontaneous breaking of the electroweak gauge group to the electric charge symmetry, $SU(3)_L\otimes U(1)_X\rightarrow U(1)_{QED}$, we allow the neutral components of the triplets to develop vacuum expectation value (VEV), $\langle \chi^0\rangle =v_\chi$, $\langle \rho^0\rangle = v_\rho$,  $\langle \eta^0\rangle = v_\eta$. We can understand this breaking as a two-step process, the first one produced by $v_\chi$, breaking $SU(3)_L\otimes U(1)_X$ to $SU(2)_L\otimes U(1)_Y$ symmetry, while $v_\eta$ and/or $v_\rho$ break $SU(2)_L\otimes U(1)_Y$ to the electric charge symmetry, $U(1)_{QED}$. We present in the Appendix the scalar potential from which we obtain the scalar's mass eigenvalues and  eigenstates. Also, in the Appendix we show the vector boson masses and the Yukawa Lagrangian involving the quarks.

It has to be said that our results are applicable not only for the two model versions we have taken into account  but also for  other similar versions with different scalar particle content as well \cite{331ss}. This is because addition/subtraction of few charged scalar particles do not impact significantly the  amplitude for the Higgs decay into two photons.

An important difference about the versions of the 3-3-1 model we deal with here is on the bounds that are imposed to their vector bosons by current available data~\cite{muonium,dion}. While these bounds can be very stringent to vector masses, $M_{U^{\pm\pm}}\geq 850$~GeV, in the minimal 3-3-1 due to muonium-antimuonium conversion and $M_{V^\pm}\geq 440$~GeV from the \emph{wrong} muon decay $\mu^- \rightarrow e^- \bar{\nu}_\mu \nu_e$, they may be alleviated in the EL331 if these are sufficiently heavy (few hundreds GeV). This is because the doubly charged vector can couple to two ordinary leptons in the minimal version, and only with an ordinary plus an exotic leptons in the EL331.
However, there are studies that can release the vector boson masses from such strong constraints even in the minimal 3-3-1 if the scalars and mixing in the leptonic sector are taken into account~\cite{VP,NM}. The lack of any signal from any model beyond SM gives us the freedom to choose these last possibilities as working tools to study the resonant decay $H\rightarrow \g\g$ at LHC in what follows.

\section{Higgs Decay into two photons}
\label{sec:decay}
\begin{figure}
\centering
\includegraphics[width=0.6\columnwidth]{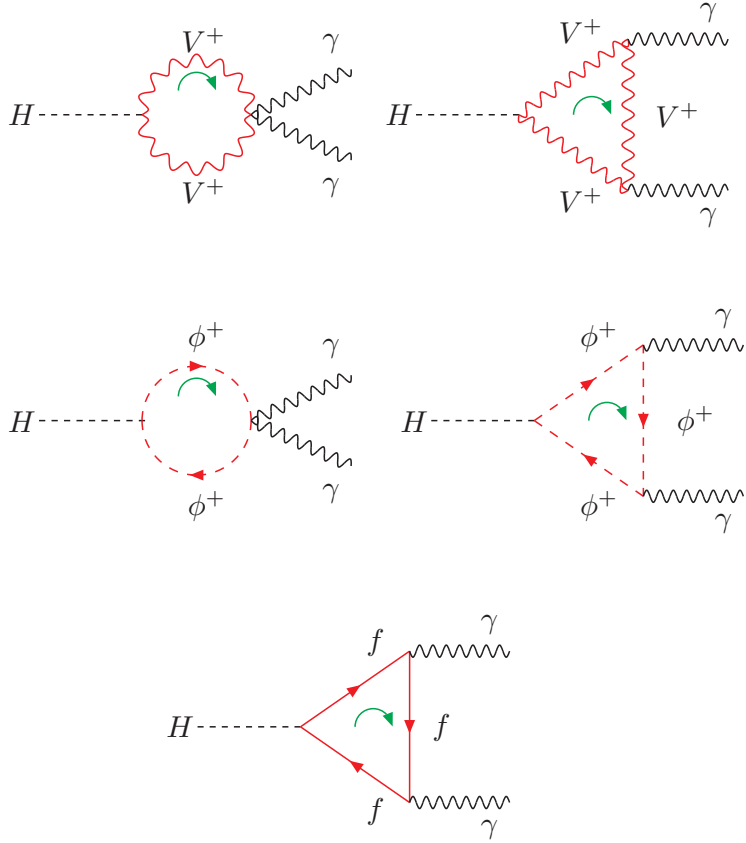}
\caption{The one-loop diagrams that contribute to the $H\rightarrow\gamma\gamma$ decay amplitude in a generic model.}
\label{fig1}
\end{figure}

The Higgs decay into two photons is possible through effective operators only (loops) and its analytical decay rate is well known for the SM~\cite{Ellis,Okun,Hunters}.
In general, the diagrams contributing to the Higgs decay amplitude into two photons are given in Fig.~\ref{fig1}, for unspecified charged vectors, scalars and fermions.
We are going to specialize to the case where the Higgs does not couple just proportionally to the particles masses. This happens in models with an extended scalar sector when the particle masses are generated by more than one VEV, which is the case of the models we are dealing with. Moreover, in these models the Higgs eigenstate is a linear combination of the neutral CP-even scalars, which may diminish somehow the Higgs-diphoton interaction strength. Following the Refs.~\cite{Ellis,Okun,Hunters}, we can write an effective Lagrangian for the Higgs-diphoton interaction by first specifying the relevant interactions of the Higgs field ($H$) with fermions ($\psi_i$), vectors ($V_{i\mu}$), and scalars ($\phi_i$) in a general fashion as
\br
{\cal L}_{int} &=& -(\sqrt{2}G_F)^\frac{1}{2}M_{\psi_i}\bar{\psi_i}\psi_i H + 2(\sqrt{2}G_F)^\frac{1}{2}m_W^2 c_{V_i} V_i^\mu V_{i\mu} H \nonumber \\
&&+ g^2 V_i^{*\mu} V_{i\mu} H^2 -2(\sqrt{2}G_F)^\frac{1}{2}M_{\phi_i}^2\phi_i^* \phi_i H - \l_i \phi_i^*\phi_i H^2\,.
\label{lint2}
\er
where $G_F$ is the Fermi constant, $m_W$ is the $W$ boson mass, and $g$, $\l_i$ the electroweak and scalars self couplings, respectively. The mass parameter, $M_{\psi_i}$, is not generally the fermion mass, as it would be the case of SM. It only means the Higgs contribution to the fermion mass in a more complex scalar sector, where the neutral scalars can mix among themselves and several VEVs may appear to engender the spontaneous symmetry breaking.  In other words, the Higgs boson coupling to the other fields is not generally proportional to their masses. Similarly, we introduce the coefficients $c_{V_i}$ for the charged vectors, so as to take account of this peculiar feature. For the present models such coefficients, under the assumptions for the VEVs ($v_\eta=v_\rho$) and the scalar fields self couplings as shown in the Appendix, are $c_{W}=1$,  $c_{U}=c_{V}=\frac{1}{2}$. The same is true for the physical charged scalars that are not present in the SM, and $M_{\phi_i}^2$ are to be treated as parameters with squared-mass dimension, not to be identified directly with their masses. In any particular model, all these parameters can be easily recognized and are well defined, as is our case in the 3-3-1 model. From the interaction Lagrangian, Eq.~(\ref{lint2}), we can obtain the following effective Lagrangian for the $H-\g-\g$ coupling,
\br
{\cal L}_{H\g\g} &=& \sum_i \frac{\alpha N_{ci} Q_i^2 F_i}{8\pi}(\sqrt{2}G_F)^\frac{1}{2} H F^{\mu\nu}F_{\mu\nu}\,,
\label{leff}
\er
where $N_{ci}$ is the color factor for the particle in the loop ($N_{ci}=1$  for singlet color fields), $Q_i$ is the ratio of the electric charge of the corresponding field to the positron one, and the form factors $F_i$ are given by,
\br
F_{\psi_i} &=& -2\tau_{\psi_i}\left[1+(1-\tau_{\psi_i})I^2\right]\frac{M_{\psi_i}}{m_{\psi_i}}\,, \nonumber \\
F_{V_i} &=& \left[2+3\tau_{V_i} +3\tau_{V_i}\left(2-\tau_{V_i}\right)I^2\right]\frac{m_{W}^2}{m_{V_i}^2}c_{V_i}\,, \nonumber \\
F_{\phi_i} &=& \left[\tau_{\phi_i}\left(1-\tau_{\phi_i}I^2\right)\right]\frac{M_{\phi_i}^2}{m_{\phi_i}^2}\,,
\label{Fs}
\er
where $m_{\psi_i}$, $m_{V_i}$,  $m_{\phi_i}$ correspond to the masses of fermionic, vector and scalar particles, respectively, with
\be
\tau_i \equiv \frac{4 m_i^2}{m_H^2}\,,
\ee
and
\be
I \equiv \left\{
\begin{array}{l}
arcsin\left(\sqrt{\frac{1}{\tau_i}}\right)\,\,\,\hbox{for} \,\,\,\tau_i\geq 1 \\
\frac{1}{2}\left[\pi+\imath \ln\left[\frac{1+\sqrt{1-\tau_i}}{1-\sqrt{1-\tau_i}}\right]\right]\,\,\,\hbox{for} \,\,\,\tau_i\leq 1
\end{array}\right.
\ee
Finally, the Higgs decay rate into two photons can be obtained,
\be
\Gamma_{H\g\g} = \frac{\alpha^2m_H^3 G_F}{128\sqrt{2}\pi^3}\left|\sum_i N_{ci} Q_i^2 F_i \right|^2\,.
\label{largura}
\ee
It is opportune to recall that for a SM Higgs-like sector, a kind of decoupling violation occurs in the $H\rightarrow\g\g$ decay rate since the Higgs couples proportionally to the particles masses and heavier particles develop a more important contribution than lighter ones.
In multi-Higgs scenarios, that is not completely true. As one can see from Eq.~(\ref{Fs}), the vector boson couplings carry a suppression factor $m_W^2c_{V_i}/m_V^2$. This mass ratio suppression appears due to factoring out the Fermi constant, namely, $(\sqrt{2}G_F)^\frac{1}{2}$, in Eq.~(\ref{lint2}), diminishing the contribution from vectors  much heavier than the $W^\pm$. In the limiting case of heavy particles, compared to the Higgs mass, Eq.~(\ref{Fs}) can be approximated to,
\br
F_{\psi_i} &\approx & -\frac{4}{3}\frac{M_{\psi_i}}{m_{\psi_i}}\,, \nonumber \\
F_{V_i} &\approx& 7 \frac{m_{W}^2}{m_{V_i}^2}c_{V_i}\,, \nonumber \\
F_{\phi_i} &\approx&-\frac{1}{3}\frac{M_{\phi_i}^2}{m_{\phi_i}^2}\,.
\label{Fs2}
\er
From this result, we can contemplate the role of new vector bosons in models beyond SM. In the 3-3-1 models, the minimal and the EL331, there are two additional singly charged, $V^\pm$, and two doubly charged, $U^{\pm\pm}$, vector bosons, potentially increasing the signal of diphoton Higgs decay. Other 3-3-1 models also possess a couple of charged vectors, but the models studied here have a further factor of enhancement which is the double charge of one of these vectors, since the electric charge appears in Eq.~(\ref{largura}) to the fourth power. We can also observe that although scalars contribute destructively, their importance is generally marginal since for heavy scalars the form factor in Eq.~(\ref{Fs2}) is only a fraction of fermion and vector boson ones. Therefore, scalars  would compete with the vector  bosons  only for very light scalars and/or for too many physical scalars in the spectrum. While the latter option is not going to be the case for the 3-3-1 models here investigated, the former could be a possible window for the charged scalar sector of the theory.  We postpone this investigation for a future work though, here we restrict our analysis to portions of the parameter space where the charged scalars are not too light. Nevertheless, we will see in the next section that a barely visible effect of lighter scalars shows up already for the Higgs and charged vector's masses accessible for the CERN LHC running at 7 TeV of center-of-mass energy.

\section{Phenomenological analysis}
\label{phenomenology}

The lightest neutral CP-even Higgs boson of the 3-3-1 models, which we call just Higgs boson for short, has the same tree-level couplings to the SM weak bosons $Z$ and $W$, leptons and quarks, as the SM Higgs boson. Furthermore, the Higgs boson in these models lack couplings to the new heavy quarks and/or leptons (these turn out to be singlets under the $SU(2)_L$ symmetry). For these reasons, the main production mode for such a Higgs is the usual gluon fusion process~\cite{djouadi1}. As a matter of fact, the cross sections from all production modes are expected to be the same size as the SM Higgs.

Moreover,  Higgs boson here should have the same branching fractions to those SM particles to which it couples at tree-level as the SM Higgs. The loop-induced decays, however,  receive contributions from the new scalars and vector bosons of the model. This is indeed the case for the decays $H\to\g\g$ and $H\to Z\g$, but not for $H\to gg$, once the Higgs here does not couple to the new heavy quarks. As a consequence, the total and partial widths of the 3-3-1 Higgs boson are nearly identical to the SM Higgs ones except for the $\g\g$ and $Z\g$ channels.

Any expected deviations from the SM case concerning the Higgs boson production and decay at hadron colliders are due solely to the branching ratio into two photons and $Z$ plus a photon, and not from the production cross sections or decays into the other channels.
This interesting feature allows us, for instance, to easily apply the exclusion limits from the Tevatron~\cite{Tevatron} recent results to the 3-3-1 models parameters space, although the present amount of data is just barely enough to constrain the model. Until its closure, however, the Tevatron has the potential to exclude a larger region of the 3-3-1 masses and couplings.

As we discussed in the previous section, the indirect bounds on the charged vector masses from muonium-antimuonium conversion and muon decays~\cite{muonium,dion} can be evaded when the effect of the destructive interference between the scalar and vector states are properly taken into account. Even in those cases where the interference is not relevant, mixing effects in the leptonic sector could weaken the couplings responsible for muonium-antimuonium conversion and muon decays. Moreover, present bounds from direct searches of a $W^\prime$~\cite{wplhc} boson may also be evaded in the case of the minimal 3-3-1 model.

\begin{figure}
\centering
\includegraphics[scale=0.65]{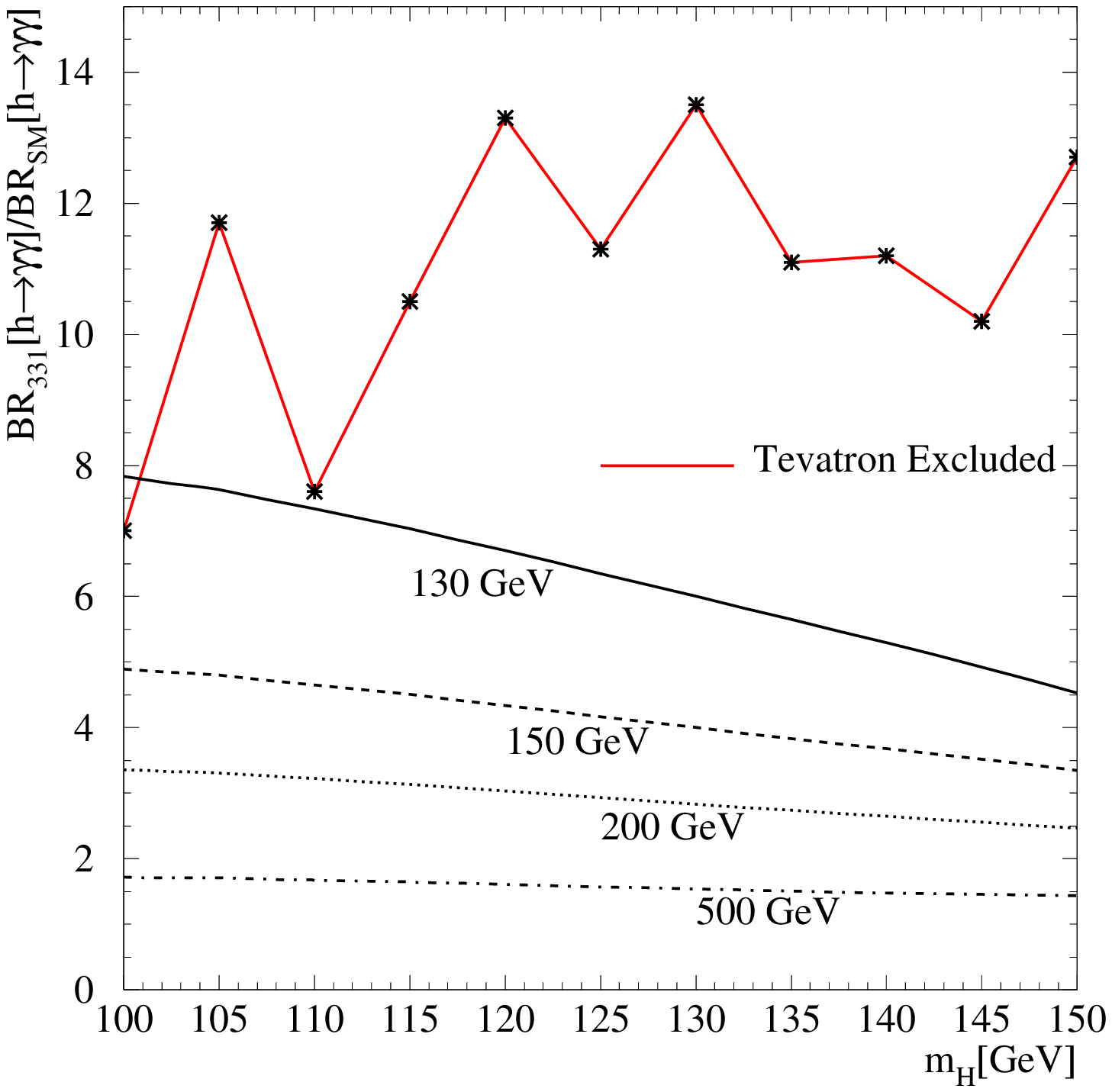}
\includegraphics[scale=0.65]{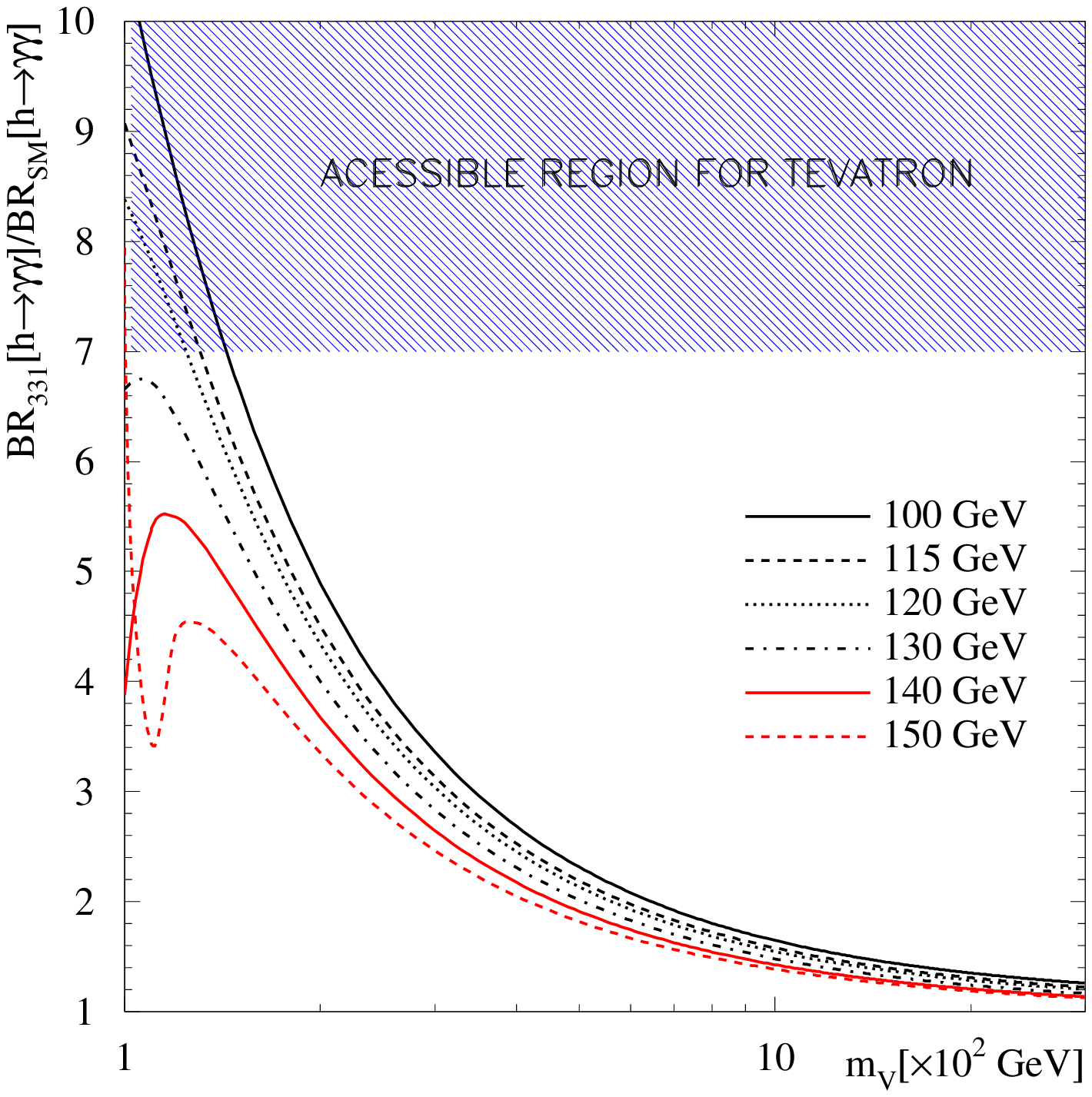}
\caption{At the upper panel, the ratio between the branching fractions from the 3-3-1 model and the SM of a light Higgs boson into photons as a function of the Higgs mass. The current Tevatron exclusion limit is above the red line in this panel. At the lower panel, the same ratio but as a function of the charged vectors mass. We also show in the lower panel the accessible region for Tevatron with the current statistics.}
\label{fig:bratios}
\end{figure}

We show in Fig.~\ref{fig:bratios} the ratio between the branching fractions of the Higgs boson into a pair of photons from the 3-3-1 model and the SM. The branching ratio of $H\to\g\g$ was calculated by adapting the {\texttt Hdecay} program from the formulas
\br
Br_{331}(H\to\g\g) & = & \frac{\Gamma_{331}(H\to\g\g)}{\Gamma_{331,total}} \\
\Gamma_{331,total} & = & \Gamma_{SM,total}-\Gamma_{SM}(H\to\g\g)+\Gamma_{331}(H\to\g\g)
\label{eq:br331}
\er
where $\Gamma_{331}(H\to\g\g)$ was calculated at one loop-level from the contributions depicted in Fig.~\ref{fig1}, and all the other partial widths calculated from {\texttt Hdecay}, including QCD and EW corrections as implemented in the program. The QCD two-loop corrections to $H\to \g\g$ partial width are very small for light Higgs masses ($m_H < 200\gev$)~\cite{djouadi1} and are not accounted for in {\texttt Hdecay}, so our LO calculation has the same accuracy as the SM Higgs computed by the program.

At the upper panel of Fig.~\ref{fig:bratios}, we show the ratio as a function of the Higgs mass for some charged vector masses and plot altogether the 95\% C.L. exclusion limit from Tevatron~\cite{Tevatron}. For a 120 GeV Higgs, we see that a light vector mass results in a larger ratio as expected. For $m_V = 200$~GeV, the enhancement compared to the SM case is around a factor 3. The decoupling behavior as the new vector boson mass increases is also evident. Note that a $10\invfb$ amount of data will start to probe effectively the 3-3-1 parameters space.
At the lower panel of Fig.~\ref{fig:bratios}, we display the ratio as function of the new charged vector boson mass for six different Higgs boson masses. The decoupling is more visible in this plot. We also note that the Tevatron  already excludes very light new vector boson masses.

Notice that in the lower panel of Fig.~\ref{fig:bratios} a small dip, shows up for intermediate Higgs masses ($\sim 150$~GeV). At the left of the dip, the partial width into two photons increases substantially. This corresponds to the regime where the Higgs can produce a pair of on-shell charged scalars, $m_H > 2 m_{h^+}$, as can be demonstrated writing the charged scalar masses as a function of the neutral Higgs and the vector mass. At the right of the dip, the partial width increases as the mass of the charged scalars diminishes reducing the overall effect of the destructive interference between the scalar and the vector loop amplitudes. However, as the masses of vectors increase, the decoupling behavior takes place, bringing the partial width closer to the SM value.

The enhanced decay rate into photons is a very distinctive signature of the 3-3-1 models that can already be probed at the low luminosity runs of the CERN LHC running at 7 TeV of center-of-mass energy~\cite{ATLAS}. Moreover, the large mass spectrum of the model affects significantly the branching ratios of the light Higgs boson into photons, as we discussed above.

In order to show the potential of the 7 TeV LHC to probe the class of 3-3-1 models we performed a simulation of the process
\be
pp\to H\to \g\g + X
\label{eq:pphaa}
\ee
and the main irreducible and reducible backgrounds.

The Higgs boson production in the gluon fusion mode was simulated at parton level and normalized to the NLO QCD plus electroweak corrections using the {\texttt HIGLU} program~\cite{higlu}. The CTEQ6~\cite{cteq} structure functions were used for all computations fixing the renormalization and the factorization scales to the $\g\g$ invariant mass, $\mu_R=\mu_F=m_{\g\g}$.

The most important irreducible background is the double photon production $pp\to\g\g$. This contribution was computed including NLO QCD corrections to the direct production, the one-loop box $gg\to\g\g$ contribution, and the contributions from single and double photon bremssthralung $pp\to q(q^\prime)\g\g$ processes using the {\texttt DIPHOX}~\cite{diphox} package. Concerning  the fragmentation processes, a photon was considered isolated from any other activity in the acceptance region if the transverse energy $E_T$ deposited inside a cone of radius $\Delta R_{\g X}=0.4$ around the photon's direction not exceeding $4\gev$.

The reducible backgrounds: $pp\to \g j$, $pp\to jj$, and $pp\to e^+e^-$ are as important as the irreducible one and were computed taking into account the NLO QCD corrections.

We assume that a light quark or gluon jet can fake a single isolated photon with probability $P_{j\to\g}=2\times 10^{-4}$~\cite{ATLAS2}. Because of the large rate for prompt $\g j$ production, this process represents the second largest background and accounts for $\sim 30$\% of all fake events. In order to simulate this background as accurately as possible we used the {\texttt JETPHOX}~\cite{jetphox} program with NLO QCD corrections plus the single photon fragmentation contribution (Bremsstrahlung). The photon isolation criteria was the same as the double photon background case.

The $pp\to jj$ events were generated with the help of {\texttt ALPGEN}~\cite{alpgen} and a K-factor of 1.3 was included to approximately take the NLO QCD contributions into account. High-energy electrons and positrons irradiate photons by interacting with the inner detector. Such photons may mimic our signal and constitute a reducible background from electrons and positrons produced via the Drell-Yan process. Previous studies from the ATLAS collaboration indicate that there is a $P_{e\to\g}=0.112$ probability per electron to fake a prompt photon. We generated $e^+e^-$ events from Drell-Yan process using {\texttt MadEvent}~\cite{madevent} and applied that probability to simulate this reducible background.

For the signal and all backgrounds, we impose the following experimentally driven set of acceptance cuts on each particle identified as a photon
\be
p_T > 40\; \hbox{GeV},\;\;\;\;\; |\eta| < 2.47,\;\;\;\;\; \Delta R_{ij} > 0.4
\label{eq:cuts1}
\ee
An identification efficiency $\varepsilon_{\g}=90$\% for true photons was also taken into account which corresponds to a 1/5000 rejection factor against jet fakes~\cite{ATLAS2}. The detector resolution effects were accounted for by smearing the photons energies, but not their directions, as done in Ref.~\cite{ATLAS}.

After applying the acceptance cuts, we search for a Higgs resonance in a $5$~GeV window around the Higgs masses in the $\g\g$ invariant mass distribution, $M_{\g\g}$
\be
|M_{\g\g}-m_H| < 2.5\; \hbox{GeV}
\label{eq:cuts2}
\ee

The impact of the acceptance cuts plus the search cut on signals and backgrounds are displayed in Table~\ref{tab:cuts}.
\begin{table}
\begin{center}
\begin{tabular}{c|cccccc}
\hline
\hline
$m_H$~[GeV]  & 100 & 110 & 120 & 130 & 140 & 150\\
\hline
Signal & 27.2 & 35.0 & 37.3 & 33.1 & 24.7 & 14.7 \\
\hline
$\g\g$ & 284.2 & 232.0 & 195.6 & 162.3 & 130.9 & 100.8 \\
$\g j$ & 144.5 & 111.9 & 83.9 & 65.1 & 47.7 & 37.6 \\
$jj$ & 32.7 & 27.4 & 23.9 & 20.1 & 16.1 & 14.1 \\
$e^+e^-$ & 0.10 & $\;\; <0.1$ & $\;\; <0.1$ & $\; <0.01$ & $\; <0.01$ & $\; <0.01$ \\
\hline
Total Bckg & 461.5 & 371.3 & 303.4 & 247.5 & 194.7 & 152.5 \\
\hline
\hline
\end{tabular}
\end{center}
\caption{Signal and backgrounds cross sections after cuts and efficiencies in fb. The signals were calculated for $200$~GeV charged vectors and the scalar parameters discussed in the text.}
\label{tab:cuts}
\end{table}
The signal cross sections displayed in the table were calculated for $m_{V^+}=m_{U^{++}}=200\gev$ charged vectors. The masses of the charged scalars vary as a function of the neutral Higgs mass from  $m_{H^{++}}=m_{h_2^+}=282\gev$ and $m_{h_1^+}=362.5\gev$ for $m_H=100\gev$ to $m_{H^{++}}=m_{h_2^+}=261.2\gev$ and $m_{h_1^+}=344.8\gev$ for $m_H=150\gev$ which span the Higgs masses displayed in the table.

\begin{figure}
\centering
\includegraphics[scale=0.9]{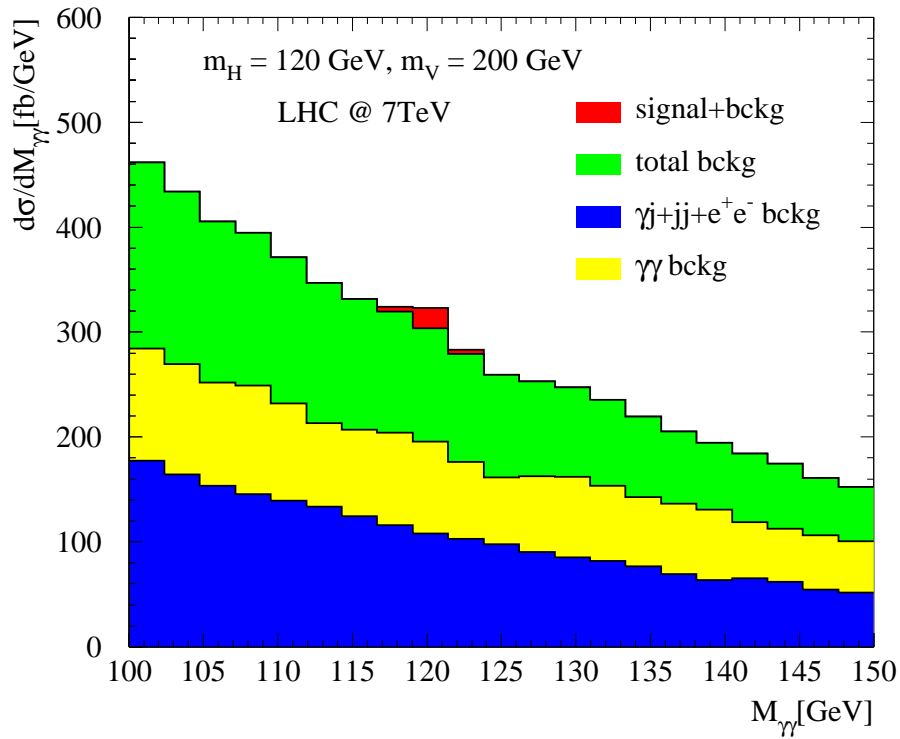}
\caption{The photon pair invariant mass, $M_{\g\g}$, distribution for a $120$~GeV Higgs boson and all the relevant backgrounds. The set of parameters are the same as those described in the text.}
\label{fig:maa}
\end{figure}

The backgrounds agree reasonably well with those quoted in Ref.~\cite{ATLAS} noting that we require two photons with at least $40$~GeV of transverse momentum which is a harder requirement than that found in Ref.~\cite{ATLAS}.
In the Fig.~\ref{fig:maa}, we show the photon pair invariant mass distribution for a $120$~GeV Higgs and all the relevant backgrounds in the $100 < m_H < 150$~GeV range, where the Higgs into photons signal is more promising. The small excess in the $120\gev$ can be made more visible after the background subtraction in a sideband analysis as has been demonstrated in the experimental studies~\cite{ATLAS}. In this case, a more careful treatment of statistical and systematic errors in the background estimation is necessary to assess the correct signal significance.

The statistical significance of the signal over the backgrounds was computed from
\be
S = \sqrt{2\left[(s+b)\ln\left(1+\frac{s}{b}\right)-s\right]}
\label{eq:stat}
\ee
where $s={\cal L}\times\sigma_S$ is the number of signal events and $b={\cal L}\times\sigma_B$ the number of background events for a given integrated luminosity ${\cal L}$.

\begin{figure}
\centering
\includegraphics[scale=0.55]{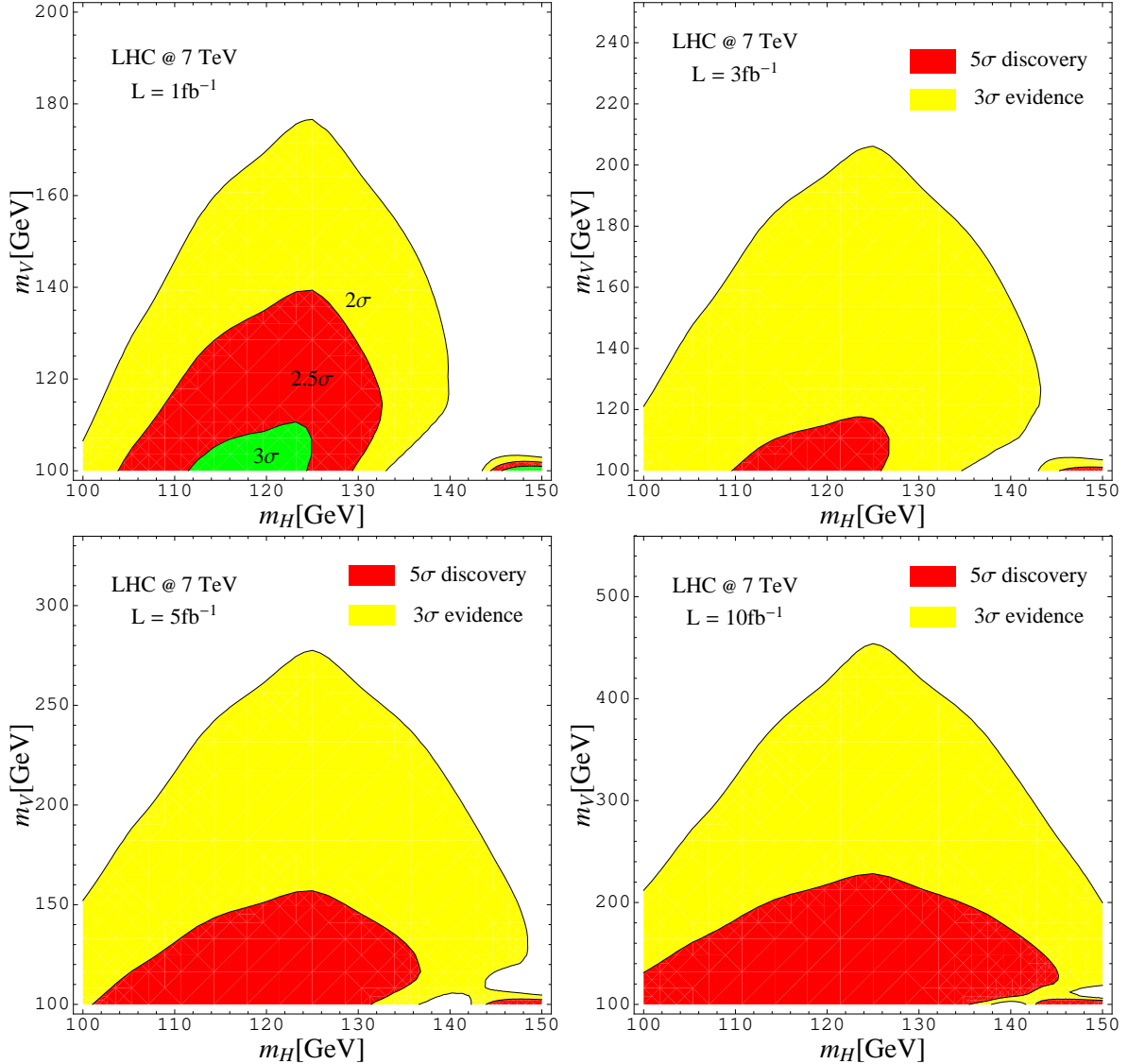}
\caption{At the left upper panel, we display the portions of the $m_H\times m_V$ plane where a level of statistical significance similar to those reported at~\cite{ATLAS2011,ATLASnew,CMS} can be reached. In the other panels, the yellow regions represent a $3\sigma$ evidence whereas the red regions represent a $5\sigma$ discovery.}
\label{fig:disc}
\end{figure}

We show in Fig.~\ref{fig:disc} the regions in the $m_H\times m_V$ plane where a $S=3\sigma$ evidence and a $S=5\sigma$ discovery is possible for an integrated luminosity of $1$, $3$, $5$, and $10\; \hbox{fb}^{-1}$ at the 7~TeV LHC. At the left upper panel, we display the $L=1\;\invfb$ case which represents the reach for the current amount of data at the LHC. We see that a light $100-150$~GeV charged vector can account for the $\sim 2\sigma$ excess reported by the combined ATLAS and CMS collaborations~\cite{ATLAS2011,ATLASnew,CMS}.

In particular, the limits reported in the dedicated analysis by the ATLAS Collaboration~\cite{ATLASnew} for the $H\to \g\g$ channel can be already translated to small portions of the parameters space of the 3-3-1 models studied in this work. Their results show that $\sigma_{new}/\sigma_{SM}$ from $2$ to $5.8$ can be excluded at 95\% C.L. in the Higgs mass range $110$--$150\gev$. This is in the ballpark of our results but a complete simulation including all the experimental issues will be necessary.

It is important to point out that a $\sim 150\gev$ charged spin-1 boson can be, in principle, a candidate to account for the resonance in the $Wjj$ channel reported by the CDF collaboration~\cite{wjjcdf}. The EL331 model with exotic leptons could be a natural model to explain such excess but the new charged vector boson $V^+$ couples to exotic quarks only. As far as we know, it remains an open question if the minimal version can be a viable model too by adjusting the VEV configurations in such a way that the $V^+$ couplings to the SM leptons are suppressed. Anyway, we believe that if the forthcoming data confirms that excess, it will require additional  studies to further investigate the 3-3-1 models, and extended gauge sector models in general, against the Tevatron data concerning the $Wjj$ excess. Also, it has to be said that the 3-3-1 model can explain  the top forward/backward asymmetry recently measured at the Tevatron, as was shown in Ref.~\cite{topfb331}.

By the end of this year the integrated luminosity expected at the LHC may reach the $5\invfb$ mark and with these data a, $\sim 120\gev$ Higgs boson and charged vectors up to $150\gev$ can be discovered as can be seen in Fig.~\ref{fig:disc}. If the run is extended to accumulate $10\invfb$, the reach increases to $\sim 200\gev$ for $120\gev$ Higgs bosons.

\begin{figure}
\centering
\includegraphics[scale=0.8]{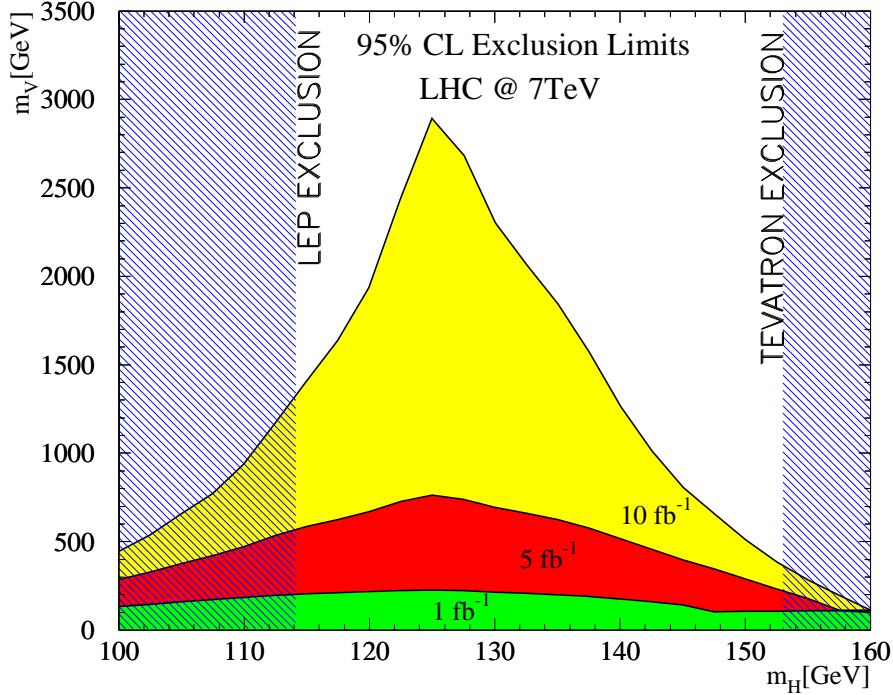}
\caption{The 95\% C.L. exclusion limits in the $m_V\times m_H$ plane for $1$, $5$, and $10\invfb$ of integrated luminosity at the 7 TeV LHC. The shaded bands show the LEP and the Tevatron excluded regions for a SM Higgs boson.}
\label{fig:excl}
\end{figure}

On the other hand if no statistically significant excess is observed above the SM backgrounds the LHC has a great potential to exclude large portions of the space of parameters. Fig.~(\ref{fig:excl}) shows the portion of the $m_H\times m_V$ plane excluded at 95\% Confidence Level for $1$, $5$, and $10\invfb$. If we interpret the actual excess as being just a background fluctuation, then the current data already exclude the light charged vectors region independent of the value of the Higgs boson mass. As the branching ratio of the light Higgs boson into $WW$ and $ZZ$ is the same as in the SM, the LEP~\cite{LEP} and the Tevatron~\cite{tevatronH} limits apply to our case and the regions excluded in these experiments are also shown in the Fig.~\ref{fig:excl}.

With $5\invfb$, the LHC starts to perform better than the indirect experiments in the task to probe the 3-3-1 models, and until the closure of the first phase of the LHC running, $1\; \hbox{TeV}$ charged vectors at least can be excluded for $110\lesssim m_H\lesssim 140\gev$ with $10\invfb$. This is far beyond the reach of any indirect search experiment.

The prospects for the 14~TeV LHC are much more promising but we postpone this study to a future work~\cite{future}, where a combination with other production modes and decay channels will be performed using a multivariate analysis. Meanwhile we emphasize the importance of a deeper investigation of the 3-3-1 models, and extended gauge sectors in general, to potentially explain the $Wjj$ excess and the top forward/backward asymmetry reported by the Tevatron collaborations in connection to the possible excess seen in the $H\to\g\g$ channel at the LHC.

\section{Conclusions}
\label{sec:conclusions}

In this work, we investigated the potential of the 7~TeV CERN LHC to search for a light neutral Higgs boson from 3-3-1 models in the $pp\to H\to \g\g$ channel.

The presence of additional singly and doubly charged vector bosons in the $H\to \g\g$ loop enhances the Higgs boson branching ratio into a pair of photons compared to the SM case. The enhancement may reach a factor $10$ or more depending on the new vector bosons masses. However, due to the striking capabilities of the CERN LHC to detect a light Higgs boson, smaller enhancement factors may be reached by the experiment, and large portions of the parameters space of these models can be probed with up to $10\invfb$ of integrated luminosities. For example, if $m_H = 120\gev$, then charged vectors with up to $\sim 400\gev$ can be discovered at the $5\sigma$ statistical level even for very heavy scalar masses.

The potential of the LHC to exclude portions of the parameters' space if no statistically significant signal is observed is far beyond the current indirect experiments.
In the absence of any signal, TeV scale masses can be excluded at 95\%  confidence level with $10\invfb$, and charged vectors with masses of several hundred GeV with up to $5\invfb$.

A small excess of $\sim 2\sigma$ has been reported by the LHC collaborations~\cite{ATLAS2011,ATLASnew,CMS} in the $H\to \g\g$ channel with $1\invfb$ of data. Such a level of statistical excess is expected for a Higgs boson from 3-3-1 models for charged vector bosons masses around $150\gev$. Interestingly these masses would correspond to a $Z^\prime$ vector boson mass in these models which can explain  the observations at the Tevatron of top forward/backward asymmetry as well \cite{topfb331}.  Moreover, if a construction is possible where the new charged vector bosons are leptophobic, then a $\sim 150\gev$ $V^+$ could explain the $Wjj$ excess at Tevatron and the $\sim 2\sigma$ excess in the Higgs search at the LHC.
We believe these coincidences are strong enough to motivate further studies for the 3-3-1 and other similar extended-gauge sector models.

\acknowledgments
This work was supported by Conselho Nacional de Pesquisa e
Desenvolvimento Cient\'{i}fico - CNPq, Coordena\c{c}\~ao de Aperfei\c{c}oamento Pessoal de N\'ivel Superior - CAPES, and Funda\c{c}\~ao de Amparo \`a Pesquisa do Estado de S\~ao Paulo - FAPESP.

\appendix
\section*{Appendix A}

\subsection{The mass eigenstates}

Considering the gauge and Lorentz invariance, we can write down the most general renormalizable scalar potential for this model,
\begin{eqnarray} V(\eta,\rho,\chi)&=&\mu_\chi^2 \chi^2 +\mu_\eta^2\eta^2
+\mu_\rho^2\rho^2+\lambda_1\eta^4 +\lambda_2\rho^4
+\lambda_3\chi^4+ \nonumber \\
&&\lambda_4(\eta^{\dagger}\eta)(\rho^{\dagger}\rho)
+\lambda_5(\chi^{\dagger}\chi)(\eta^{\dagger}\eta)+\lambda_6(\chi^{\dagger}\chi)(\rho^{\dagger}\rho)
+ \nonumber \\
&&\lambda_7(\eta^{\dagger}\rho)(\rho^{\dagger}\eta)
+\lambda_8(\chi^{\dagger}\eta)(\eta^{\dagger}\chi)+\lambda_9(\chi^{\dagger}\rho)(\rho^{\dagger}\chi)
  -\frac{f}{2}\epsilon_{ijk}\eta_i\rho_j\chi_k + \mbox{H.c}.
\label{potential}
\end{eqnarray}

In order to achieve spontaneous symmetry breaking, we assume that the neutral scalars ($\eta^0 ,\, \rho^0 ,\, \chi^{0})$ develop the following VEV,
\begin{eqnarray}
 \eta^0 , \rho^0 , \chi^{0} \rightarrow  (v_{\eta ,\rho ,\chi}
+R_{ \eta ,\rho ,\chi} +iI_{\eta ,\rho ,\chi})\,,
\label{vacua}
\end{eqnarray}

We can then obtain the following minimum conditions (tadpole conditions) from the potential in Eq.~(\ref{potential}),
\begin{equation}
 \mu^2_\chi +2 \lambda_3 v^2_{\chi} +\lambda_5 v^2  + \lambda_6 v^2- a \frac{v^2}{2}=0\,,
\end{equation}
\begin{equation}
\mu^2_\eta +2 \lambda_1 v^2 + \lambda_4 v^2 + \lambda_5 v_{\chi}^{2} -a\frac{v_{\chi}^2}{2} =0\,,
\nonumber
\end{equation}
\begin{equation}
\mu^2_\rho +2 \lambda_2 v^2 + \lambda_4 v^2+ \lambda_6 v^2_\chi-a \frac{v_{\chi} }{2} =0\,,
\end{equation}
where we have defined the mass parameter $f$ in the potential as $f\equiv -a v_\chi$ and $v_\rho = v_\eta \equiv v$ (the SM VEV is given by $v_w^2 = v_\rho^2+v_\eta^2 \approx 246^2$~GeV$^2$).

We can then write the CP-even scalars' mass matrix in the basis  ($R_\eta ,\, R_\rho ,\, R_\chi)$,
\begin{eqnarray}
\left(\begin{array}{ccc}
R_\chi & R_\eta & R_\rho
\end{array}\right)\left(\begin{array}{ccc}
\frac{a v^2}{2} + 4 \lambda_3 v_\chi^3 & -\frac{a v v_\chi}{2} + 2 \lambda_5 v v_\chi & -\frac{a v v_\chi}{2} +2 v v_\chi \lambda_6 \\
 -\frac{a v v_\chi}{2} + 2 \lambda_5 v v_\chi  & \frac{a v_\chi^2}{2}+4\lambda_1 v^2 & -\frac{a v_\chi^2}{2} + 2 \lambda_4 v^2 \\
-\frac{a v v_\chi}{2} + 2 v v_\chi \lambda_6 & -\frac{a v_\chi^2}{2} + 2 \lambda_4 v^2 & \frac{a v_\chi^2}{2} + 4 \lambda_2 v^2 \\
\end{array}\right)\left(\begin{array}{c}
R_\chi \\
R_\eta \\
R_\rho \\
\end{array}\right)
\end{eqnarray}

We assume that $\lambda_6=\lambda_5=\frac{a}{4}$ in order to obtain a simple analytical solution for the mass eigenstates,
\begin{eqnarray}
S_1 & = & R_\chi,\nonumber \\
S_2  & = & \frac{1}{\sqrt{2}}( R_\rho - R_\eta) \nonumber \\
H  & = & \frac{1}{ \sqrt{2} }( R_\rho + R_\eta).
\end{eqnarray}
and their respective eigenvalues,
\begin{eqnarray}
\frac{m_{S_1}^2}{2} & = & \frac{a v^2}{2} + 4 \lambda_3 v_\chi^2,\nonumber \\
\frac{m_{S_2}^2}{2} & = & 1/2 \left(a v_\chi^2 + 4 v^2 (\lambda_1 + \lambda_2) + \sqrt{16 v^4(\lambda_1 - \lambda_2)^2 +(a v_\chi^2 - 4 v^2 \lambda_4)^2} \right),\nonumber \\
\frac{m_H^2}{2} & = & 1/2 \left(a v_\chi^2 + 4 v^2 (\lambda_1 + \lambda_2) - \sqrt{16 v^4(\lambda_1 - \lambda_2)^2 +(a v_\chi^2 - 4 v^2 \lambda_4)^2} \right),
\end{eqnarray}
Notice that despite the appearance of $v_\chi$ in the Higgs mass, an expansion of these expressions for $v_\chi >> v$ shows that the Higgs mass depends only on powers of $v/v_\chi$, being the lightest mass of the neutral scalars in this model.

For the CP odd scalars, the mass matrix in the basis $(I_{\chi}\,,\,I_\eta\,,\,I_\rho)$ is written as,
\begin{eqnarray}
\left(\begin{array}{ccc}
I_\chi & I_\eta & I_\rho
\end{array}\right)\left(\begin{array}{ccc}
\frac{a v^2}{2} & \frac{a v v_\chi}{2} & \frac{a v v_\chi}{2} \\
 \frac{a v v_\chi}{2} & \frac{a v_\chi^2}{2} & \frac{a v_\chi^2}{2} \\
\frac{a v v_\chi}{2} & \frac{a v_\chi^2}{2} & \frac{a v_\chi^2}{2} \\
\end{array}\right)\left(\begin{array}{c}
I_\chi \\
I_\eta \\
I_\rho \\
\end{array}\right)\,,
\end{eqnarray}
whose eigenstates are given by,
\begin{eqnarray}
G_1 & = & \frac{1}{ \sqrt{ 1+\frac{v^2}{v_\chi^2} } } (- I_\chi+  \frac{v}{v_\chi} I_\rho ),\nonumber \\
G_2 & = & \frac{1}{\sqrt{2}} \left[ \left(-\frac{v_\chi}{v} + \frac{v_\chi}{ v(1+\frac{v^2}{v_\chi^2} )}\right)I_\chi + I_\eta - \frac{1}{1+\frac{v^2}{v_\chi^2} }I_\rho \right],\nonumber \\
P_1 & = & \frac{1}{\sqrt{2+\frac{v^2}{v_\chi^2} }}\left(\frac{v}{v_\chi} I_\chi + I_\eta + I_\rho \right).
\end{eqnarray}
%
%with the defintion,
%
%\br
%A\equiv \sqrt{1+\frac{\left(a^2 v_\chi^4-8a\l_4 v^2 v_\chi^2
%+ 16\l_4^2 v^4\right)}{\left(a v_\chi^2-4\l_4 v^2\right)^2}}
%\er
%
Here, $G_{1}$ and $G_{2}$ correspond to Goldstone bosons, eaten by the $Z$ and $Z^\prime$ vector bosons, and $P_1$ is a massive CP-odd scalar that remains in the spectrum and whose mass is
\begin{eqnarray}
\frac{ m^{2}_{P_{1}}}{2} &=& \frac{1}{2} (a v^2 + 2 a v_\chi^2)
\end{eqnarray}

Concerning the doubly charged scalars, we obtain the mass matrix in the basis $(\chi^{\pm\pm}_2 \rho^{\pm\pm}_2)$,
\begin{eqnarray}
\left(\begin{array}{cc}
\chi^{--}_2 & \rho^{--}_2
\end{array}\right)\left(\begin{array}{cc}
\frac{a v^2}{2}+\lambda_9 v^2 & \frac{a v v_\chi}{2}+\lambda_9 v v_\chi \\
 \frac{a v v_\chi}{2}+\lambda_9 v v_\chi & \frac{a v_\chi^2}{2}+\lambda_9 v_\chi^2 \\
\end{array}\right)\left(\begin{array}{c}
\chi^{++}_2 \\
\rho^{++}_2 \\
\end{array}\right),
\end{eqnarray}
with the following eigenvectors, 
\begin{eqnarray}
G^{\pm\pm}_1 & = & \frac{1}{  \sqrt{(1+\frac{v_\chi^2}{v^2})}  }( -\frac{v_\chi}{v}\chi^{\pm\pm}_2 + \rho^{\pm\pm}_2)\,, \nonumber \\
H^{\pm\pm} & = &\frac{1}{  \sqrt{(1+\frac{v^2}{v_\chi^2})}  }( \frac{v}{v_\chi}\chi^{\pm\pm}_2 + \rho^{\pm\pm}_2)\,.
\label{vetoresDCS}
\end{eqnarray}
The $G^{\pm\pm}_1$ is a Goldstone boson, eaten by the doubly charged vector boson $U^{\pm\pm}$ while the remaining doubly charged scalar, $H^{\pm\pm}$, has mass,
\begin{equation}
m^{2}_{H^{\pm\pm} } = \frac{1}{2} ( av^2 + a v_\chi^2 + 2 \lambda_9 v^2 + 2 \lambda_9 v_\chi^2)
\end{equation}

Similarly, for the singly charged scalars in the basis $(\eta^{\pm}_1 \rho^{\pm}_1)$,
\begin{eqnarray}
\left(\begin{array}{cc}
\eta^{-}_1 & \rho^{-}_1
\end{array}\right)\left(\begin{array}{cc}
\frac{a v_\chi^2}{2} + \lambda_7 v^2 & \frac{a v_\chi^2}{2} + \lambda_7 v^2 \\
\frac{a v_\chi^2}{2} + \lambda_7 v^2 & \frac{a v_\chi^2}{2} + \lambda_7 v^2 \\
\end{array}\right)\left(\begin{array}{c}
\eta^{+}_1 \\
\rho^{+}_1  \\
\end{array}\right)\,
\end{eqnarray}
with mass eigenstates,
\begin{eqnarray}
G^{\pm}_1 & = & \frac{1}{\sqrt{2}}( \rho_1^{\pm} - \eta_1^{\pm} )\,, \nonumber \\
h_1^\pm & = & \frac{1}{\sqrt{2}}( \rho_1^{\pm} + \eta_1^{\pm} )\,,
\label{vetoresSCS}
\end{eqnarray}
where $G^{\pm}_1$ is a Goldstone boson, eaten by the $W^\pm$ vector boson, while $h^{\pm}_1$ remains in the spectrum with mass given by,
\begin{equation}
m^{2}_{ h^{\pm}_1 } = av_\chi^2 + 2\lambda_7 v^2
\end{equation}

Finally, in the basis $(\chi^{-}_1 \eta^{-}_2)$, we have the mass matrix,
\begin{eqnarray}
\left(\begin{array}{cc}
\chi^{-}_1 & \eta^{-}_2
\end{array}\right)\left(\begin{array}{cc}
\frac{a v^2}{2} + \lambda_8 v^2 & \frac{a v v_\chi}{2} + \lambda_8 v v_\chi \\
 \frac{a v v_\chi}{2} + \lambda_8 v v_\chi & \frac{a v_\chi^2}{2} + \lambda_8 v_\chi^2 \\
\end{array}\right)\left(\begin{array}{c}
\chi^{+}_1 \\
\eta^{+}_2  \\
\end{array}\right)\,,
\end{eqnarray}
whose eigenvectors are,
\begin{eqnarray}
G^{\pm}_2 & = & \frac{1}{\sqrt{ 1+\frac{v_\chi^2}{v^2} }}( -\frac{v_\chi}{v}\chi^{\pm}_1 + \eta^{\pm}_2 )\,, \nonumber \\
h^{\pm}_2 & = &\frac{1}{\sqrt{ 1+\frac{v^2}{v_\chi^2} }}( \frac{v}{v_\chi}\chi^{\pm}_1 + \eta^{\pm}_2 ) \,,
\label{vetoresSCS2}
\end{eqnarray}
where $G^{\pm}_2$ is a Goldstone boson, eaten by the $V^{\pm}$ vector boson, while the second charged massive scalar is $h_2^-$, with mass,
\begin{equation}
m^{2}_{ h^{\pm}_2 } = \frac{1}{2}(av^2 + a v_\chi^2 + 2 v^2 \lambda_8 + 2 v_\chi^2 \lambda_8)\,.
\end{equation}

The Yukawa Lagrangian involving the quarks can be written as,
\br
-{\cal L}_Y &=&  \l_1 {\bar Q_{1_L}}\chi J_{1_R} + \l_{mn} {\bar Q_{m_L}}\chi^* j_{n_R} +\l^\prime_{1i}{\bar Q_{1_L}}\rho d_{i_R}
\nonumber \\
&+& \l^\prime_{ni}{\bar Q_{n_L}}\rho^* u_{i_R} + \l^{\prime\prime}_{1i}{\bar Q_{1_L}}\eta u_{i_R} + \l^{\prime\prime}_{ni}{\bar Q_{n_L}}\eta^* d_{i_R} + h.c.
\label{yukawa}
\er
The Yukawa Lagrangian in Eq.~(\ref{yukawa}) provides the correct masses of the quarks.

As for the charged vector bosons the mass eigenstates are defined as:
\be
W^\pm = -\frac{W_1\mp \imath W_2}{\sqrt{2}}\,,\,\,\,\,\,\,V^\pm = -\frac{W_4\pm \imath W_5}{\sqrt{2}}\,,\,\,\,\,\,\,U^{\pm\pm} = -\frac{W_6\pm \imath W_7}{\sqrt{2}}\,
\ee
and their respective mass eigenvalues,
\be
m_W^2 = \frac{g^2}{2}v^2\,,\,\,\,\,\,\,m_V^2 = m_U^2 = \frac{g^2}{4}(v^2+v_\chi^2)\,.
\ee
%

%%%%%%%%%%%%%%%%%%%%%%%%%%%%%%%%%%%%%%%%%%%%%%

%%%%%%%%%%%%%%%%%%%%%%%%%%%%%%%%%%%%%%%%%%%%%%%%%%%%%%%%%%%%%%%%%%%%%%%%%%%%%%%%%%%%%%%%%%%

\end{document}